# Dosimetric Comparison of Passive Scattering and Active Scanning Proton Therapy Techniques Using GATE Simulation

A. Asadi[1], A. Akhavanallaf[2], S. A. Hosseini[1], H. Zaidi[2,3,4,5]

[1]Department of Energy Engineering, Sharif University of Technology, Tehran, Iran, Zip code: 8639-11365

[2]Division of Nuclear Medicine and Molecular Imaging, Geneva University Hospital, CH-1211 Geneva 4, Switzerland

[3]Geneva University Neurocenter, Geneva University, CH-1205 Geneva, Switzerland

[4]Department of Nuclear Medicine and Molecular Imaging, University of Groningen, University Medical Center Groningen, 9700 RB Groningen, Netherlands

[5]Department of Nuclear Medicine, University of Southern Denmark, DK-500, Odense, Denmark

**Corresponding Authors:**

**Short running title:** passive scanning versus active scanning proton therapy

**Abstract**

In this study, two proton beam delivery designs, i.e. passive scattering proton therapy (PSPT) and pencil beam scanning (PBS), were quantitatively compared in terms of dosimetric indices. The GATE Monte Carlo (MC) particle transport code was used to simulate the proton beam system; and the developed simulation engines were benchmarked with respect to the experimental measurements. A water phantom was used to simulate system energy parameters using a set of depth-dose data in the energy range of 120-235 MeV. To compare the performance of PSPT against PBS, multiple dosimetric parameters including Full Width at Half Maximum (FWHM), peak position, range, peak-to-entrance dose ratio, and dose volume histogram have been analyzed under the same conditions. Furthermore, the clinical test cases introduced by AAPM TG-119 were simulated in both beam delivery modes to compare the relevant clinical values obtained from Dose Volume Histogram (DVH) analysis. The parametric comparison in the water phantom between the two techniques revealed that the value of peak-to-entrance dose ratio in PSPT is considerably higher than that from PBS by a factor of 8%. In addition, the FWHM of the lateral beam profile in PSPT was increased by a factor of 7% compared to the corresponding value obtained from PBS model. TG-119 phantom simulations showed that the difference of PTV mean dose between PBS and PSPT techniques are up to 2.9 % while the difference of max dose to organ at risks (OARs) exceeds 33%. The results demonstrated that the active scanning proton therapy systems was superior in adapting to the target volume, better dose painting, and lower out-of-field dose compared to passive scattering design.

## 1. Introduction

Proton therapy (PT) due to their excellent dose distribution can significantly reduce the absorbed dose by the patient′s body relative to the photon beams [1]. PT was first initiated in 1954 and is currently going as part of modern radiation therapy technologies in many developed countries [2]. In PT, there are two main techniques of irradiation, namely active scanning or Pencil Beam Scanning (PBS) and Passive Scattering Proton Therapy (PSPT). Passive scattering facilities are composed of specific mechanical devices in the particle trajectory to shape the beam to the tumor volume using particle−matter interactions. The scattering interaction spread the originally Gaussian-distributed beam to shape a wide homogeneous beam to the tumor using patient-specific collimators. A rotating wheel of varying thickness, called range modulator, is used to generate a uniform Spread-Out Bragg Peak (SOBP). Lastly, proton beam travel through a longitudinal compensator, a specific scatterer device, specifically drilled for each field and each patient, to achieve the last conformal shaping of the beam just before the skin of the patient. Nowadays, proton beam delivery is switching from the use of passively scattered mode to pencil beam scanning owing to the feasibility of more conformal dose to tumors and higher dose rate in PBS compared to PSPT  ADDIN EN.CITE [3].  In addition, pencil beam technique can reduce the need for specific mechanical hardware required in PSPT. Another advantage of the active scanning technique is related to the lower level of secondary particles, mostly neutrons that are generated from the interaction of primary protons with multiple scattering components. In PBS, multiple magnets in x and y directions, based on the charge of particles, are used to drift the beam and scanning the target volume spot by spot with a 3D narrow pencil beam [4, 5]. There are some studies in biological dose comparison between active and passive scanning proton therapy techniques in the cellular level ADDIN EN.CITE [3, 6, 7]. Gridley et al. ADDIN EN.CITE [7] compared cell response to active scanning and passive beam delivery techniques. They reported lower dose values at the entrance region, a sharper fall-off and a higher dose rate for active scanning mode. While they did not find any considerable difference in relative biological effectiveness (RBE) between two methods. Michaelidesova et al. ADDIN EN.CITE [3] reported that between two techniques, there is neither considerable beam quality difference from MC simulation, nor statistically significant difference in biological endpoints. Nomura et al. [6] reported a higher Linear Energy Transfer (LET) at distal region in active scanning mode compared to passive scattering method. To the best of our knowledge, the current study is the first report that systematically compares physical doses between two modes of proton therapy. In this work, multiple dosimetric criteria such as FWHM, peak position, range, peak-to-entrance dose ratio, etc have been analyzed. Furthermore, clinically relevant dose parameters have been compared through TG-119 clinical test cases.

## 2. Material and method

In this work, two main proton therapy techniques, i.e. PBS and PSPT, have been simulated using Monte Carlo GATE code. Shanghai Advanced Proton Therapy (SAPT) facility, a synchrotron-based active scanning proton therapy system, was simulated. Fig 1. depicts the geometrical characteristics of SIEMENS IONTRIS system at SAPT. In this system proton beams are extracted from synchrotron and drifted to the nozzle, by using the paired scanning magnets in horizontal (X) and vertical (Y) directions. Proton beam spot is moved around the isocenter with energies between 70-235 MeV [8]. The advantage of this technique against scattering-based technique is that a range shifter is not required to shape the beam to the tumor volume, because the synchrotron accelerate the protons slowly and conform the tumor dose in the dimension lateral to the beam [8]. The system characteristics are listed in Table 1. Dose delivered to the phantom is monitored in real time by using two parallel-plate ionization chambers. Spot size and beam optic are measured by using the position detectors. Unlike discrete scan mode (pixel scan), IONTRIS provided a continuous beam scan mode (raster scan).

Table 1. The characteristics of SAPT proton therapy system obtained from [8].

| Item | value |
|---|---|
| Energy (MeV) | 70.0-235.0 |
| Field size (cm$^2$) | 30.0×40.0 |
| Scanning magnet x to isocenter distance (cm) | 287.0 |
| Scanning magnet y to isocenter distance (cm) | 247.0 |
| Nozzle to isocenter distance (cm) | 40.0 |
| Scan speed in x (cm/ms) | 2.0 |
| Scan speed in y (cm/ms) | 0.5 |
| Dose rate (Gy/min) | 2.0 |

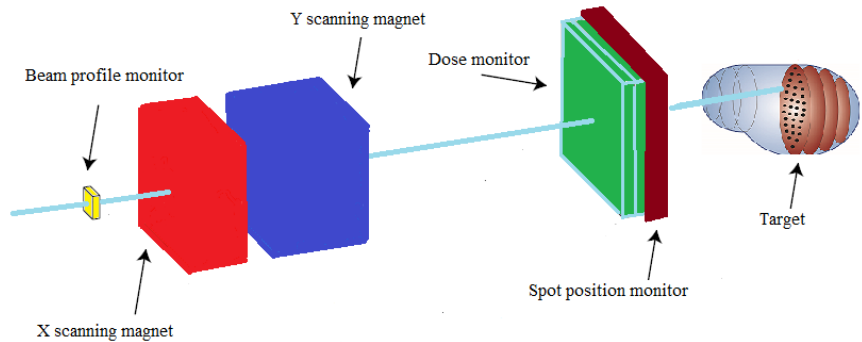

Figure 1. A schematic view of SAPT nozzle.

Passive scattering system geometry was obtained from [9]. In this design, proton beam passes through the vacuum window, first scatterer, first monitor, range modulator wheel, second scatterer, second monitor, collimator, and the size-changeable snout (brass tube). The first scatterer is set for spreading the beam laterally and the range modulator wheel spread the beam longitudinally. Two jaws can provide a size-changeable treatment field, while the aperture can control the lateral conformity of the beam.

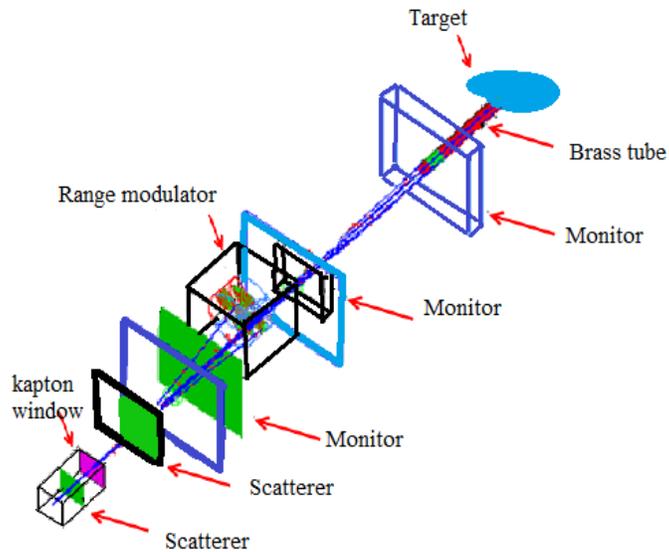

Figure 2 A schematic view of passive scattering system including scatterer, i.e. range modulator, collimator and compensator, along with monitoring components.

The strategy reported by Grevillot et al. [10] has been adopted for PBS system simulation. a complete treatment plan in PENCIL BEAM GATE SOURCE TPS including two main components, source description and plan description has been defined [10, 11]. In PBS simulation, the source description file defines the beam delivery system with a set of the polynomial equations which allows calculation of the optical and energy properties of each pencil beam at the nozzle output as a function of beam energy. It consists of eight equations; where, two equations describe energy properties and the rest of the equations describe optical properties. The plan description file contains one or more fields described by gantry angle and a set of pencil beams.

In PSPT design, the simulated geometry consists of a nozzle and a water phantom (Fig. 2) [9]. The system characteristics including the nozzle components and materials, are listed in Table 2. All the components are simulated in mm accuracy. The circular and semicircular geometries have been used to simulate the two final diaphragms. For the proton beam, the mean primary energy was set on 212.2 MeV, with a Gaussian distribution of 3.5%. In the simulation, proton beam, before entering to the nozzle, passes from 10 cm air, and after the exit window, passes from 57 mm air, just before entering to the water phantom.

Table 2. Specifications of simulated PSPT nozzle.

| Component | Material | Length (mm) | Outer radius (mm) | Inner radius (mm) |
|---|---|---|---|---|
| Range shifter tube | Brass | 254 | 82.6 | 76.2 |
| First collimator | Brass | 10 | 76.2 | 6.4 |
| Range shifter | Lexan | A | 62.5 | -- |
| Second collimator | Brass | 10 | 57.2 | 12.7 |
| Monitor chamber tube | Brass | 681 | 57.2 | 51.4 |
| Vacuum tube | Aluminum | 0.25 | 254 | -- |
| Taper tube | Brass | 203 | 44.3 | 38.5 |
| Circular aperture | Brass | 161 | 44.3 | b |
| Half-Circular aperture | Brass | 9.5 | 20 | 12 |
| Monitor chamber plate | Brass | 9.5 | 20 | 13.5 |

a: is the range shifter length e and varies in the simulation.
b: Internal and external dimensions vary depending on the therapeutic.
Brass (8.49 g/cm³) is composed of 61.5% Copper, 35.2% Zinc and 3.3% lead; Lexan density: 1.2gr/cm³

In this study, two independent Monte Carlo-based simulators for PBS and PSPT have been developed and benchmarked against experimental measurements reported in the literature. GATE code, a multipurpose Monte Carlo code based on the libraries of the Geant4 toolkit, has been used in this simulation (version 8.2 and QGSP-BERT-EMZ physical list) [12]). The physical list selected for proton transport simulation is QGSP-BERT-EMZ.

## Clinical phantom study

The clinical test suite recommended by TG-119 includes structures for prostate, head and neck (H&N), and C-shape cases. The prostate phantom, uses the CTV, PTV, rectum, and bladder; the head & neck case, includes the PTV, cord, and parotide (left and right); and the C-shape phantom uses a PTV, and a core structure. The PTV in C-shape phantom wrapped around a core, whose outer surface is 5 mm from the inner surface of the PTV [13].

## Plan description

For a given radiation field, there are two main objectives: maximizing the uniform dose at the target and minimizing the non-target dose [14]. Inverse planning algorithm proposed by Sánchez et al. [15] applied for spot and beam selection. The objective function and target in this optimization was set according to the TG-119 report [16] and other references ADDIN EN.CITE [15, 17, 18]. The accuracy of the dose volume histogram obtained from GATE simulation was benchmarked against the results reported by Sánchez et al [15]. Hence, clinically relevant dosimetric parameters were compared in terms of mean relative error using Eq. (1):

$$\text{MRE} = (\frac{\text{Di} - \text{Di}'}{\text{Di}}) \times 100 \qquad (1)$$

where, $\text{Di}$ represents the dose parameters from our simulation and $\text{Di}'$ represents the reference data. For the TG-119 C-shape phantom the dose goals was set base on the PTV ( $D95 = 50$ Gy, $D10 < 55$ Gy), and for core ($D10 < 10$ Gy). A single proton field was set on the target to maximize the biological effect [15]. For H&N case 70 GyRBE dose (in 35 fraction) was prescribed to be delivered to the PTV70 ($D_{20} < 55$ GyRBE, $D_{99} < 46.5$ GyRBE, and $D_{90} = 50$ GyRBE). While maximum dose goal to the OARs were restricted as following: dose to the cord (max < 50 GyRBE), brain stem (max < 54 GyRBE) and for parotide ($D_{50} < 20$ GyRBE). To minimize the biological effect on non-target volumes, two filed with 50, and 310 degree angles were set [15, 17]. In the prostate case, prescribe dose was selected 78 GyRBE (in 39 fraction) ADDIN EN.CITE [15, 18], with the goal dose to PTV ($D_5 < 83$Gy, and $D_{95} > 75.6$ Gy). Dose to rectum was set to V70$_{\text{GyRBE}} < 30$ % and V50$_{\text{GyRBE}} < 60$ %. Dose to bladder was restricted by V70$_{\text{GyRBE}} < 35$%, V50$_{\text{GyRBE}}$ < 60 %, and for femur it was defined V50$_{\text{GyRBE}} < 5$ %. Two parallel opposed field was used to minimize the dose effect on the normal tissue. During the planning, relative biological effect of protons was applied by a constant factor of 1.1.

## Quantitative analysis

Both PSPT and PBS facilities were benchmarked against experimental data in terms of depth dose curve and SOBP plan. The average point-to-point[1] difference and statistical analysis (T-test) between each series of data were calculated. When the MC simulators of two systems have been validated, they were compared under the same simulation parameters using water phantom and TG-119 test cases. The physical dose parameters including FWHM, peak location, range, and peak-to-entrance dose ratio were compared. In addition, based on the SOBP plan, the factor of conformity and penumbra were compared. Furthermore, DVH-driven clinical parameters obtained from TG-119 simulation were compared between two proton therapy techniques.

## 3. Results and discussion

### Experimental study

Fig. 3 shows a comparison between the simulated Integrated Depth Dose (IDD) at the energy of 161.1 MeV for PBS system, and 212.2 MeV for PSPT system, and the experimental data obtained from [8, 9]. The simulation results illustrate a good agreement with the experimental measurements. For the PBS system, the mean point-to-point difference between the simulation and measurement was 1.5% and no statistically significant difference was observed (P-value=0.008), while for PSPT system the mean point-to-point difference was calculated about 2.76% (P-

---

$$\left| \frac{di}{drefi} \right| \quad D$$

_____ Where *di* and *drefi* refers to the simulation and measurement, Δ is the step between two point, and L is the maximum range.

value=0.048).

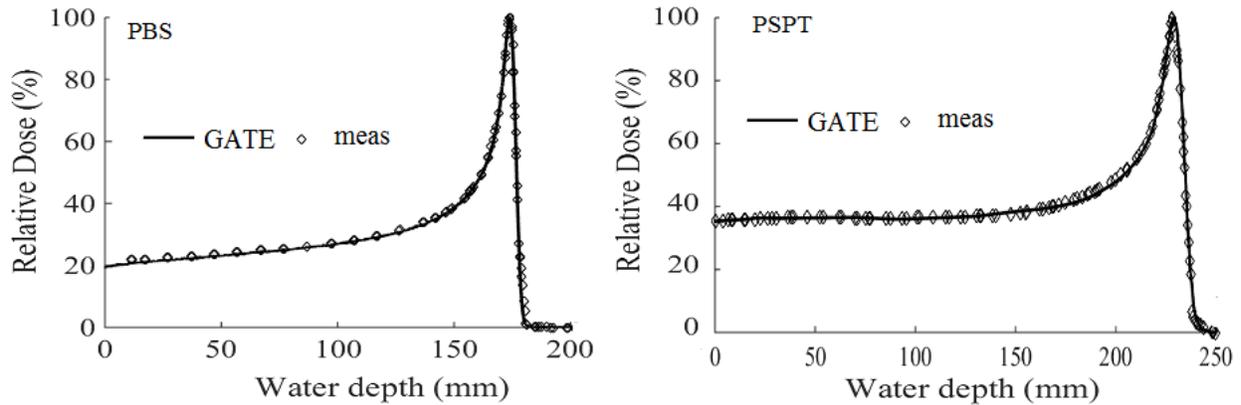

Figure 3. Comparison between the simulated and experimental data for IDD parameters, in the active scan-based method (left), and scatter-based method (right).

Fig 4. demonstrates a good agreement between the simulated SOBP and experimental data for both PBS and PSPT technique. The mean relative error for PBS was calculate about 0.97% (P-value=0.003) while for PSPT was 1.08% (P-value=0.05).

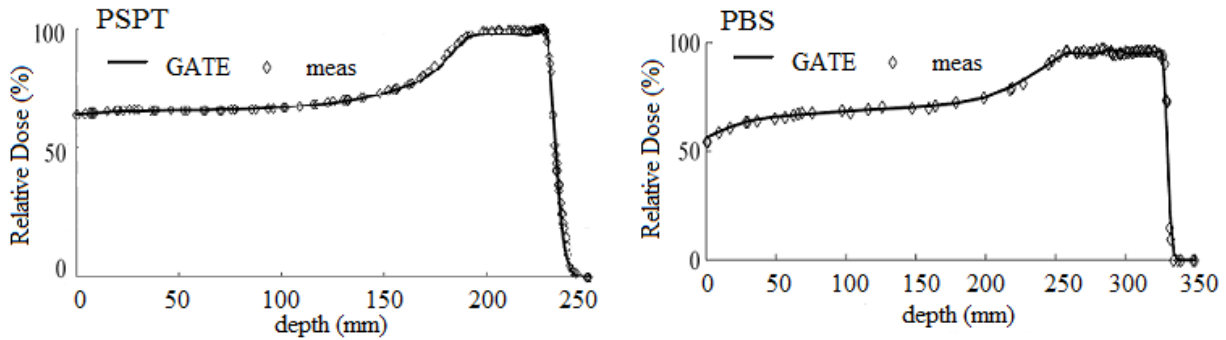

Figure 4 Comparison between the simulated and experimental data for SOBP plan, in the scatter-based method (left), and active scan-based method (right).

Fig. 5 compares the simulated Bragg peak location, range (R80), FWHM and peak-to-entrance dose ratio between PBS and PSPT beam delivery methods in different energies under the same conditions. Bragg-peak location between the two methods had a mean relative difference of 0.74% with a maximum difference of 0.99% (P-value= 0.0021). Accordingly, no significant difference on the energy discharge location has been observed between two methods. The parameter of proton beam range in the scattering is slightly longer than the active scanning method with an average difference of 1.8 mm (maximum difference 2.9 mm). In addition, comparison of FWHM parameter between scattering system versus active scanning technique show a considerable difference of about 8% (P-value=0.0022 ). The ratio of peak to entrance dose (normalized) is considerably higher with an averagy difference of 6.45 % (ranging between 1.9 - 11.11 %) in PSPT compared to PBS (P -value=0.0258).

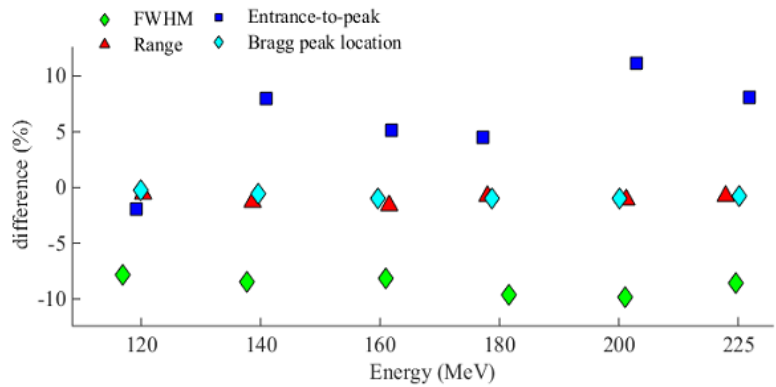

Figure 5. comparison between to mode of PT, in the term of range ( ▲ ), peak location ( ◆ ), FWHM( ◆ ), and peak-to-entrance dose ratio( ■ ).

SOBP plan and depth-dose curve in water phantom under the same conditions are compared between PBS and PSPT in Fig 6. The mean relative difference in IDD profile between two methods was calculated about 9.3% (P-value=0.0001). In PSPT mode, the slope of the fall-down in SOBP curve (normalized) is -0.03 and in PBS mode the slope of the fall-down is -0.08. The physical range extracted from SOPB curve in the passive scattering method (354.73 mm) compared to PBS mode (353.37 mm) is about 1.36 mm larger. In addition, the value of discharged dose in the modulation region for PBS mode is about 4% higher than that calculated in PSPT mode. The entrance dose in PBS mode (0.549) was lower by a factor of 13.4% compared to simulated passive scattering design

(0.623).

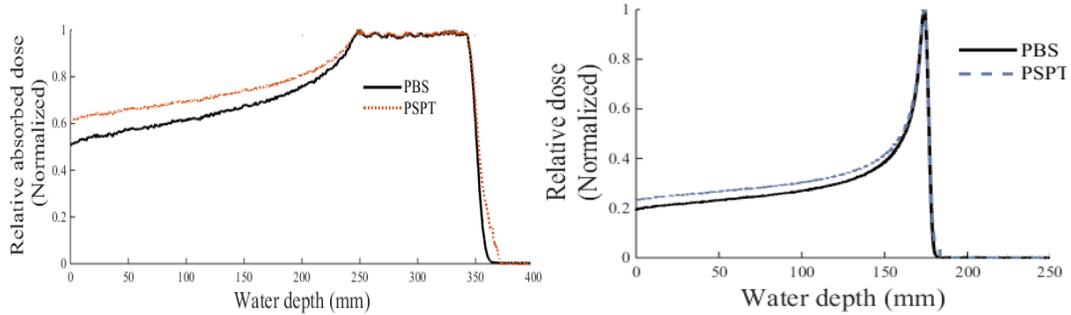

Figure 6. IDD (right) and SOBP (left) diagram obtained from simulated PBS and PSPT systems in proton energy of 179.9 MeV.

## Clinical study

Dose distribution for TG-119 phantoms in the pencil beam scanning proton therapy compared to passive scattering model along with DVH analysis for targets and OARs are illustrated in Fig 7. The clinically relevant dose parameters between two modes of proton therapy have been compared in Table 3. The results showed that the difference of PTV mean dose between PBS and PSPT techniques are up to 2.9 % while the difference of max dose to organ at risks (OARs) exceeds 33% for the spinal cord in H&N case-study.

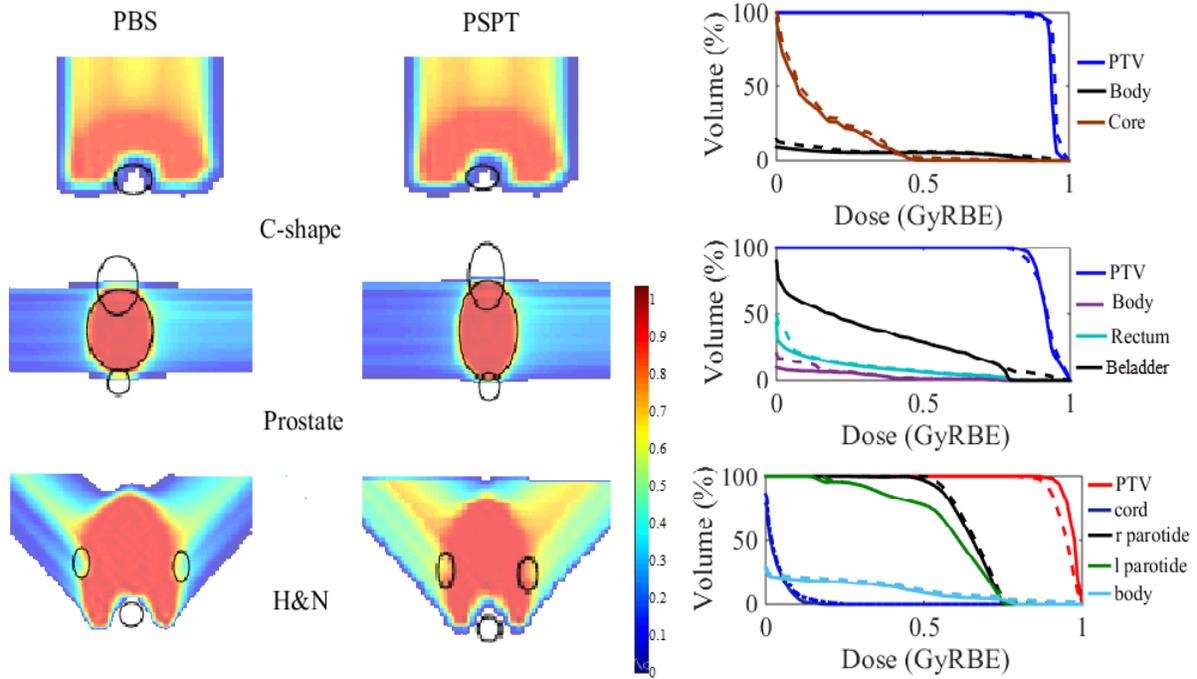

Figure 7. 3D dose distribution and DVH analysis for PBS (continues line) and PSPT (dashed line) plans.

Table 3 dose parameter obtained from dose volume histogram data

| Phantom | Organ | Indices | PSPT | PBS | Dif (%) |
|---------|-------|---------|------|-----|---------|
| C-shape | PTV | Mean dose (Gy) | 1.68 | 1.65 | 1.8 |
| | | max dose (Gy) | 1.79 | 1.77 | 1.1 |
| | | D 98 % (Gy) | 1.59 | 1.6484 | 5.8 |
| | | D 95% (Gy) | 1.65 | 1.67 | 1.2 |
| | | D 50% (Gy) | 1.715 | 1.715 | 0 |
| | | D 2 % (Gy) | 1.76 | 1.74 | 2 |
| | Core | Mean dose (Gy) | 0.29 | 0.25 | 16 |
| | | Max dose (Gy) | 1.102 | 0.917 | 20.1 |
| | | Min dose (Gy) | 0 | 0 | 0 |
| Prostate | Bladder | V80 | 7 | 0 | <100 |
| | | V75 | 10 | 0 | <100 |
| | | V70 | 21 | 20 | 4.7 |
| | | V65 | 28 | 28 | 0 |
| | Rectum | V75 | 0 | 0 | 0 |
| | | V70 | 0 | 0 | 0 |
| | | V65 | 4 | 3.8 | 5 |
| | | V60 | 10 | 10 | 0 |
| | PTV | D95 (Gy) | 74.1 | 76.44 | -3.06 |
| | | D5(Gy) | 81.07 | 81 | 0.086 |
| | | Mean dose (Gy) | 79 | 80.4 | -1.7 |
| | Cord | Max | 13.5 | 9 | 33 |
| H&N | R parotide | V32 | 69 | 65 | 5.7 |
| | L parotide | V32 | 48 | 48 | 0 |
| | PTV | D95 (Gy) | 61.8 | 63 | 1.9 |

| | 69 | 70 | 1.4 |
| D5(Gy) | 69 | 70 | 1.4 |
| Mean dose (Gy) | 67 | 69 | 2.9 |

| D5(Gy) | 69 | 70 | 1.4 |
| Mean dose (Gy) | 67 | 69 | 2.9 |

## 4. Discussion and Conclusion

Nowadays, the treatment of cancerous tumors by particles is increasing due to their outstanding features in the delivery of tailors. In this study, the main methods of beam delivery in proton therapy were simulated, and essential quantities such as peak Bragg location, range, FWHM, the ratio of dose at peak to input, and spread out Bragg peak on the target volume in a cubic water phantom were evaluated at different energies. Furthermore, the DVH analysis based on the clinical TG-119 test cases have been investigated. According to our results, some advantages in terms of physical dose parameters in active scanning model compared to passive scattering mode has been observed. However, factor of proton range and peak location were not significantly different between the two nozzle designs. A sharper slope of fall-down in SOBP curve was observed in PBS mode compared to PSPT enabling a more controllable performance to adapt dose to the target volume; where, the sharper the curve, the higher the beam control and the better the dose adjustment on the target area. FWHM in depth-dose profile was on average 8% larger in scattering system compared to active scanning mode which stems from scattering events and secondary particle generation in the trajectory of proton beam. The ratio of peak-to-entrance dose within the target volume was another investigated quantity in the present work. By comparing this quantity, we found that the inactive dispersion of the transmitters deposited in the skin and surrounding healthy tissue was significantly lower in the active scanning method. The total deposited energy in the out-of-modulation region in PSPT was about 20% higher than PBS mode (Fig 6, area under SOBP curve); which is an important advantage of PBS in clinical application. It can be seen that in the active scanning method, the tumor volume can be better covered and less stitches can be delivered to healthy tissues. Based on the DVH-driven parameters (Table 3), it can be seen that the capability of the PBS mode to reduce the dose to non-target volumes is superior to PSPT while the total dose to body contour is reduced in average about 29.9% in three TG-199 test cases. In the high dose region of bladder in prostate DVH that is quantified by V80 and V75 indices, the outperformance of PBS versus PSPT has been illustrated. In addition, better conformity of dose to PTV is observed from the fall-down slope in DVH analysis (Fig 7).

In the present study, the results of the simulation of pencil beam and scattering systems were validated against the experimental data. Therefore, we believe that the simulations performed in this study have the capability to be utilized as independent dose engine simulator of the given systems and applicable in the quality assurance process. Furthermore, under the same conditions and in a homogeneous water phantom, the quantities related to both dose and adaptability properties were investigated and the superiority of spot scanning method in both dose transfer and adaptation was investigated. According to the literature, there are some controversies regarding the distal dose of PSPT against PBS systems  ADDIN EN.CITE [3, 6, 7]; it might stem from the fact that the estimated physical/ biological dose in passive scattering proton therapy systems are dependent on the specific scattering hardwares that are implemented in the beam trajectory. In this regard, the performance of multiple scattering designs should be simulated and further compared to provide a global conclusion about the advantages and disadvantages of active scanning proton therapy technique against passive scattering mode. In the next step, it is better to do such a study in the tissue and in conditions such as the presence of implants to clarify the importance of therapeutic methods in dose delivery (and ability of PT mode to reduce the dose of secondary products).

## Acknowledgment

The authors are grateful to the research office of the Sharif University of Technology for the support of the present work.